\documentclass[12pt]{article}
\usepackage{amssymb,graphicx,graphics,epsfig}
\usepackage{lscape,graphics,amsmath}
\usepackage{latexsym,amsfonts}
\usepackage[english]{babel}
\usepackage{color}
\usepackage{cite}

\begin{document}


\begin{center}
{\large \bf NOTES ON THE INVERSE COMPTON SCATTERING}\\

\vspace{4mm}{K.A. Bornikov$^1$, I.P. Volobuev$^2$, Yu.V.
Popov$^{2,3}$}\\

\vspace{4mm}{
$^1$Physical Faculty, Lomonosov Moscow State University, Moscow, Russia\\
$^2$Skobeltsyn Institute of Nuclear Physics,
Lomonosov Moscow State University, Moscow, Russia \\
$^3$Bogoliubov Laboratory of Theoretical Physics,  Joint Institute for Nuclear Research,\\
Dubna, Moscow Region, Russia}
\end{center}

\begin{abstract}

The paper  deals with kinematic conditions for the inverse Compton
scattering  of photons by relativistic electrons and the
polarizations of the colliding particles, which affect the value
of the differential cross section of the process. A significant
influence  of the electron and photon helicity  on the value of
the cross section has been found. In the ultrarelativistic case, a
surprising effect of an almost twofold increase in the cross
section of scattering in the direction of the initial electron
momentum  has also been discovered, when the initial photon
momentum is transverse to that of the initial electron.

\end{abstract}



\section*{\large INTRODUCTION}

At the present time, the state pays a great attention to the
development of neutron and synchrotron research. ''Federal
scientific and technical program for the development of
synchrotron and neutron research and research infrastructure for
2019 --2027 years'' \cite{prog} has been announced. In the context
of our work, we would like to highlight the following topics of
this program: ''Methods of synchrotron and neutron diagnostics of
materials and nanoscale structures for promising technologies and
technical systems, including fundamentally new nature-like
component base,''\ ''Methods of synchrotron and neutron studies of
structure and dynamics of biological systems at different levels
of organization (biomolecules, macromolecular complexes, viruses,
cells),'' \, ''New technologies of electron and proton
accelerators necessary for development of new sources of
synchrotron radiation of the 4th and subsequent generations, X-ray
laser free electrons and pulsed neutron sources'',\ etc. This
program is directly related to the inverse Compton scattering.

The effect discovered by Compton 100 years ago \cite{com}
consisted in a decrease in the frequency (energy) of a photon as a
result of its scattering by a charged particle (electron) at rest.
This discovery  was of an enormous methodological importance for
quantum mechanics, since it confirmed the behavior of the photon
as a particle and formed the basic concept of wave-particle
duality in quantum mechanics. From the very beginning the Compton
effect was  aimed at studying the distribution of the electron
momentum in the targets, which originally were predominantly
solid-state ones \cite{DuMond}. Recently, it was found that
experiments on Compton scattering of photons at bound electrons
could be conducted with gas targets and slow (cold) atoms using
the COLTRIMS detector \cite{Kir}. The theory of precisely such
nonrelativistic experiments can be found, in particular, in papers
\cite{Hou, Vest23}

Much later, the attention of scientists, and especially engineers,
was attracted by the so-called inverse Compton effect, which gave
a possibility to build  relatively compact X-ray and even gamma
radiation sources \cite{survey,LEG}. The inverse Compton effect
consists in the (head-on) collision of a photon beam with a beam
of relativistic charged particles (usually electrons). The
scattered photons acquire much more energy than the initial
photons, and they can be used further, which is largely an
engineering challenge. Some aspects of Compton scattering
 at nucleons are presented in \cite{trudy,baranov68},
although, as shown below, protons are hardly appropriate to be
used for the purposes of the inverse Compton effect in the
laboratory.  The inverse Compton effect often manifests itself in
astrophysical processes involving ultrarelativistic particles
\cite{JETP64, PR68, RMP70, BBGDP90, UFN98}.

The final photons collected and collimated into a beam can be used
to investigate deep levels of heavy atoms and atomic nuclei. In
this case, it is possible to study the Compton excitation of
nuclei, various photonuclear and photoatomic processes
\cite{PhysRep}, and even to use high-energy photons in such
fantastic projects as tomography of ultrarelativistic nuclei by
photon-gluon interactions \cite{pgc}.

In this paper, we consider some aspects of the inverse Compton
effect, with particular attention to the polarization effects. We
aim to consider the kinematic conditions providing the maximum
cross section at relatively moderate photon energies, quite
achievable in laboratory lasers. An extensive theoretical material
has been accumulated during many years of research into this
elastic interaction of  photons with charged particles, which is
partly included in the textbooks. Theoretical description of this
process is carried out within the framework of quantum
electrodynamics (QED) and in the lowest order of perturbation
theory is given by two two-vertex diagrams (s- and t-channel, see
for example \cite{AB}) In this approximation, the cross section of
the process is calculated analytically. The reader can also find a
number of interesting experimental and theoretical details  in
reviews \cite{UFN54,UFN98,UFN04}.

In the paper we mainly use the atomic system of units:
$m_e=\hbar=|e|=1$. In these units, the speed of light is $c=137$
and  the fine structure constant $\alpha=1/c$, the classical
electron radius  $r_0=\alpha^2$.

\section*{\large THEORY}

\subsection*{\small COMPTON'S FORMULA FOR A MOVING TARGET}

The theory of the inverse Compton effect  is formulated mainly in
the framework of relativistic quantum electrodynamics.
Conservation of energy-momentum in the interaction of a photon
with a relativistic charged particle has the form:
$$
E_i+\omega_i=E_f+\omega_f,  \eqno (1.1)
$$
$$
\vec p_f=\vec p_i+\vec Q.  \eqno (1.2)
$$
In (1) $\vec Q=\vec k_i-\vec k_f$ is the  momentum transfer,
$E=\sqrt{p^2c^2 + m^2c^4}$ is the energy of the particle
(electron), $\omega=ck$ is the photon energy (frequency), and
$\vec k$ is its momentum. Substituting the expression for momentum
$\vec p_f$ from (1.2) into $E_f$ and using (1.1), we obtain after
simple calculations
$$
\omega_f=\omega_i\
\frac{\gamma-\cos\theta_i\sqrt{\gamma^2-1}}{\gamma-\cos\theta_f\sqrt{\gamma^2-1}+(\omega_i/mc^2)
(1-\cos\theta_{fi})}. \eqno (2)
$$
In (2) $\gamma=E_i/mc^2$, the particle momentum $\vec p_i$ is
directed along the $z$ axis,  $\theta_i$ is the angle between the
vectors $\vec p_i$  and $\vec k_i$,  $\theta_f$ is the angle
between the vectors $\vec p_i$ and $\vec k_f$, and $\theta_{fi}$
is the angle between the  vectors $\vec k_i$ and $\vec k_f$. At
last,
$\cos\theta_{fi}=\cos\theta_i\cos\theta_f+\sin\theta_i\sin\theta_f\cos\Phi$.
Next, we use the relation $\vec p=\vec v (E/c^2)$ and get $\gamma
= 1/\sqrt{1-\beta^2}$, where $\beta=v/c<1$ denotes the ratio of
the speed of the particle to the speed of light, i.e. $\gamma$ is
the standard Lorentz-factor. Finally
$$
\omega_f=\omega_i\
\frac{(1-\beta\cos\theta_i)}{(1-\beta\cos\theta_f)+\sqrt{1-\beta^2}\
(\omega_i/mc^2) (1-\cos\theta_{fi})}. \eqno (3)
$$
For $v=0$, we obtain the Compton formula for the scattering of a
photon by a  target at rest. For the inverse Compton scattering,
it is usually assumed that $\theta_i=\pi$ and
$$
\omega_f=\omega_i\
\frac{(1+\beta)}{(1-\beta\cos\theta_f)+\sqrt{1-\beta^2}\
(\omega_i/mc^2) (1+\cos\theta_{f})}. \eqno (4)
$$
The photon frequency ratio for this case is shown in Fig. 1.

If in this case the scattered photon moves forward along the axis
$z$, then $\theta_f=0$, and
$$
\omega_f=\omega_i\ \frac{(1+\beta)}{(1-\beta)+\sqrt{1-\beta^2}\
(2\omega_i/mc^2)}. \eqno (5)
$$
In the ultrarelativistic limit $\beta\to 1$, the photon energy
$\omega_f$ can formally be arbitrarily large. In this case, the
final frequency, as follows from (5), becomes independent of the
initial frequency, and
$$
\omega_f\approx \frac{mc^2}{\sqrt{1-\beta^2}}=E_i. \eqno (5.1)
$$
This formula remains valid for small deviations of the angle
$\theta_f$ from zero. Moreover, from (1.1) follows $E_f\approx
\omega_i$. It is clear that in the ultrarelativistic case we have
a large energy transfer  to the initial photon. Hence, in
particular, it follows that the use of electrons is energetically
more favorable than protons, because at the same energy of proton
and electron, the photon will receive less energy when colliding
with proton due to its larger mass.
\begin{figure}[ht!]
\centering
\includegraphics[scale=0.35]{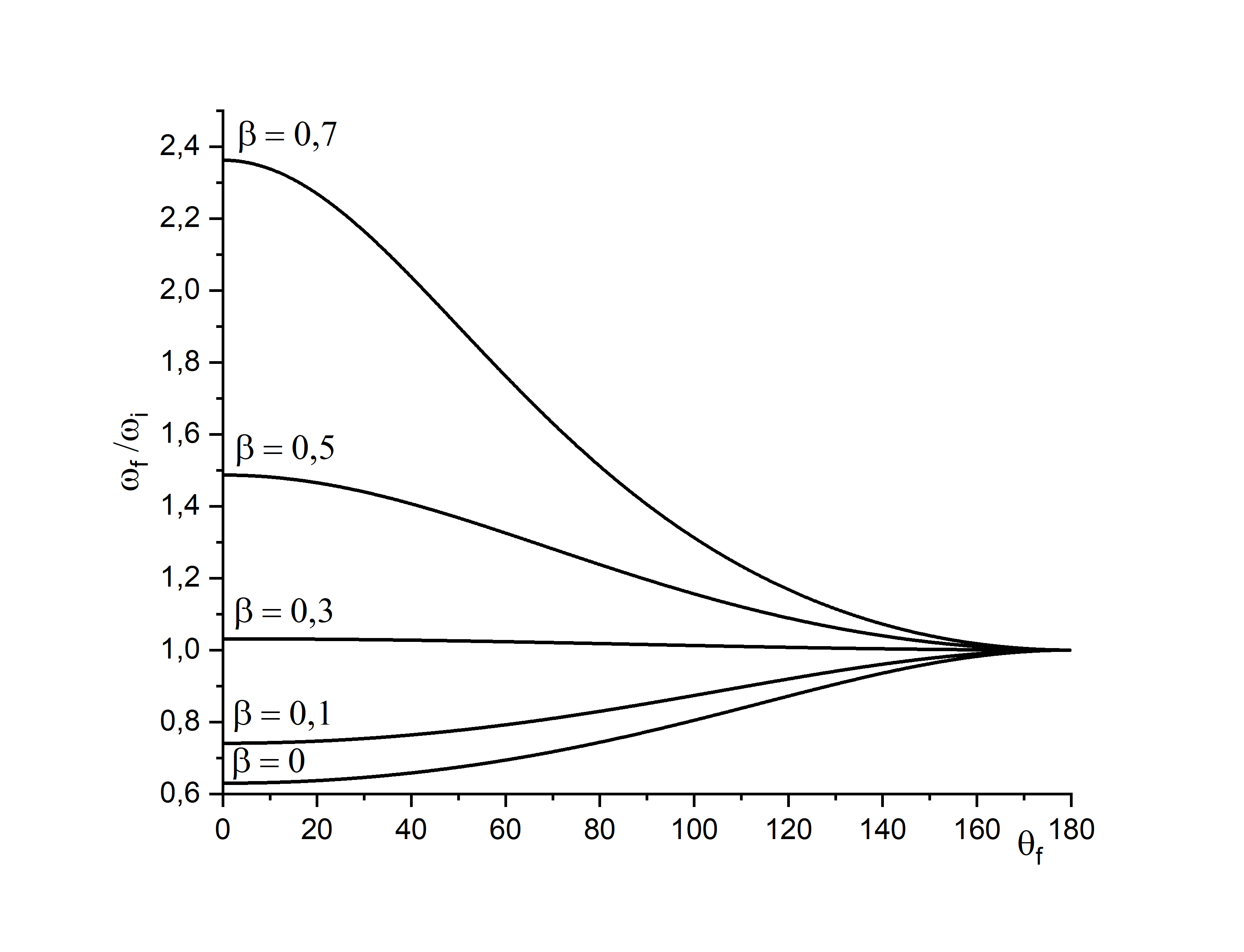}
\caption{\large The ratio of photon frequencies depending on  the
scattering angle $\theta_f$ for different values of $\beta$ (4).
The energy of the incident photon is $\omega_i=150$ keV.}
\end{figure}

\subsection*{\small DIFFERENTIAL CROSS SECTION}

The differential cross section for the scattering of the final
photon into a solid angle $\Omega_f$ for unpolarized target
particles and photons can be taken from  monograph \cite{AB} and,
when written in atomic units for an arbitrary particle mass $m$,
takes the form
$$
\frac{d\sigma}{d\Omega_f}=2r^2_0\left(\frac{\omega_f}{mc^2\kappa_1}\right)^2U_0, \eqno (6.1)
$$
where
$$
U_0=4\left(\frac{1}{\kappa_1}+\frac{1}{\kappa_2}\right)^2 -4\left(\frac{1}{\kappa_1}+\frac{1}{\kappa_2}\right) - \left(\frac {\kappa_1}{\kappa_2}+\frac {\kappa_2}{\kappa_1}\right),\eqno (6.2)
$$
$$
\kappa_1=\frac{2}{m^2c^4}(E_i\omega_i-c^2p_ik_i\cos\theta_i)= 2\left(\frac{E_i}{mc^2}\right) \left(\frac{\omega_i}{mc^2}\right)(1-\beta\cos\theta_i) ,
$$
$$
\kappa_2= - \frac{2}{m^2c^4}(E_i\omega_f-c^2p_ik_f\cos\theta_f)=-2\left(\frac{E_i}{mc^2}\right) \left(\frac{\omega_f}{mc^2}\right)(1-\beta\cos\theta_f).
$$
For the following calculations we put $\cos\theta_i=-1$ and
$\cos\theta_{if}=-\cos\theta_f$. We recall that
$$
\frac{E_i}{mc^2}=\frac{1}{\sqrt{1-\beta^2}}.
$$
and $\omega_f$ is related to $\omega_i$ by formula (3). Some
examples of the cross sections are shown in Fig. 2 (left panel).
It can be seen from the figure that, as the target speed
increases, the  graph of the cross section squeezes to small
scattering angles, which is well known and quite expected.

The formula for  cross section (6.1) has the axial symmetry about
the $z$ axis, which is related to the choice of the angle of
incidence of the photons  $\theta_i =\pi$, however, one can
consider other values of the angle of incidence. Here we calculate
cross section (6.1) averaged over the angle $\Phi$ entering
Eq.(3):
$$
\overline{\frac{d\sigma}{d\Omega_f}} = \frac{1}{2\pi r_0^2 }\ \int_0^{2\pi} \frac{d\sigma}{d\Omega_f} d\Phi,     \eqno (7)
$$
The  results of the calculations  for the angle
$\theta_i=90^\circ$ are shown in Fig. 2 (right panel). The
averaged cross sections (7) give practically constant small values
in the range of angles $\theta_f\geq 90^\circ$, and as $\beta$
increases, each subsequent curve decreases faster than the
previous one, which we also see in the left panel. An increase (a
significant on) is observed for the angles $\theta_f\sim 0$ with
increasing electron velocities. The general structure of the
curves in the figures at small scattering angles is approximately
the same, but the cross section for $\beta=0.9$ on the right panel
is about 2 times larger (!).

\begin{figure}[ht!]
\centering
\includegraphics[scale=0.35]{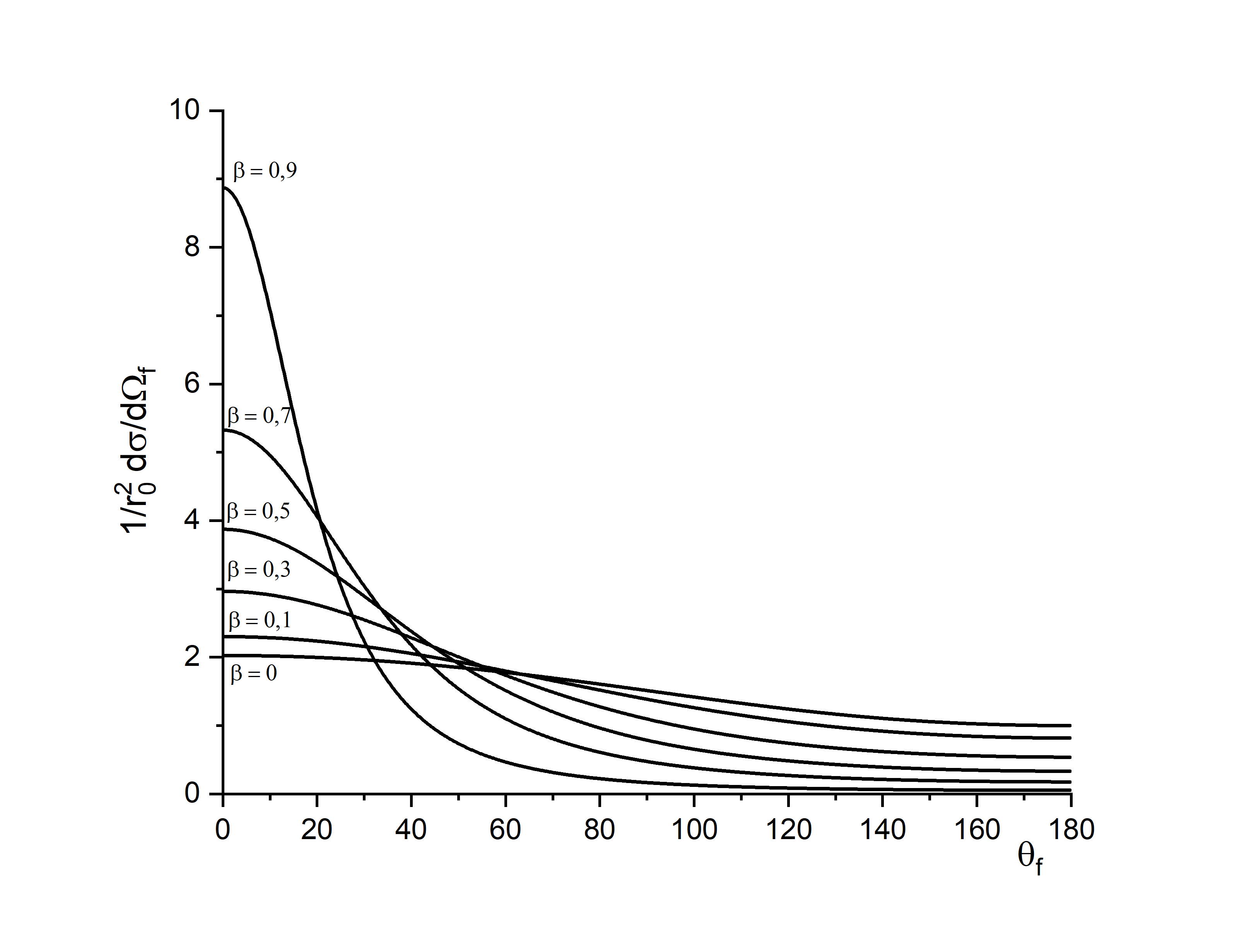}\includegraphics[scale=0.35]{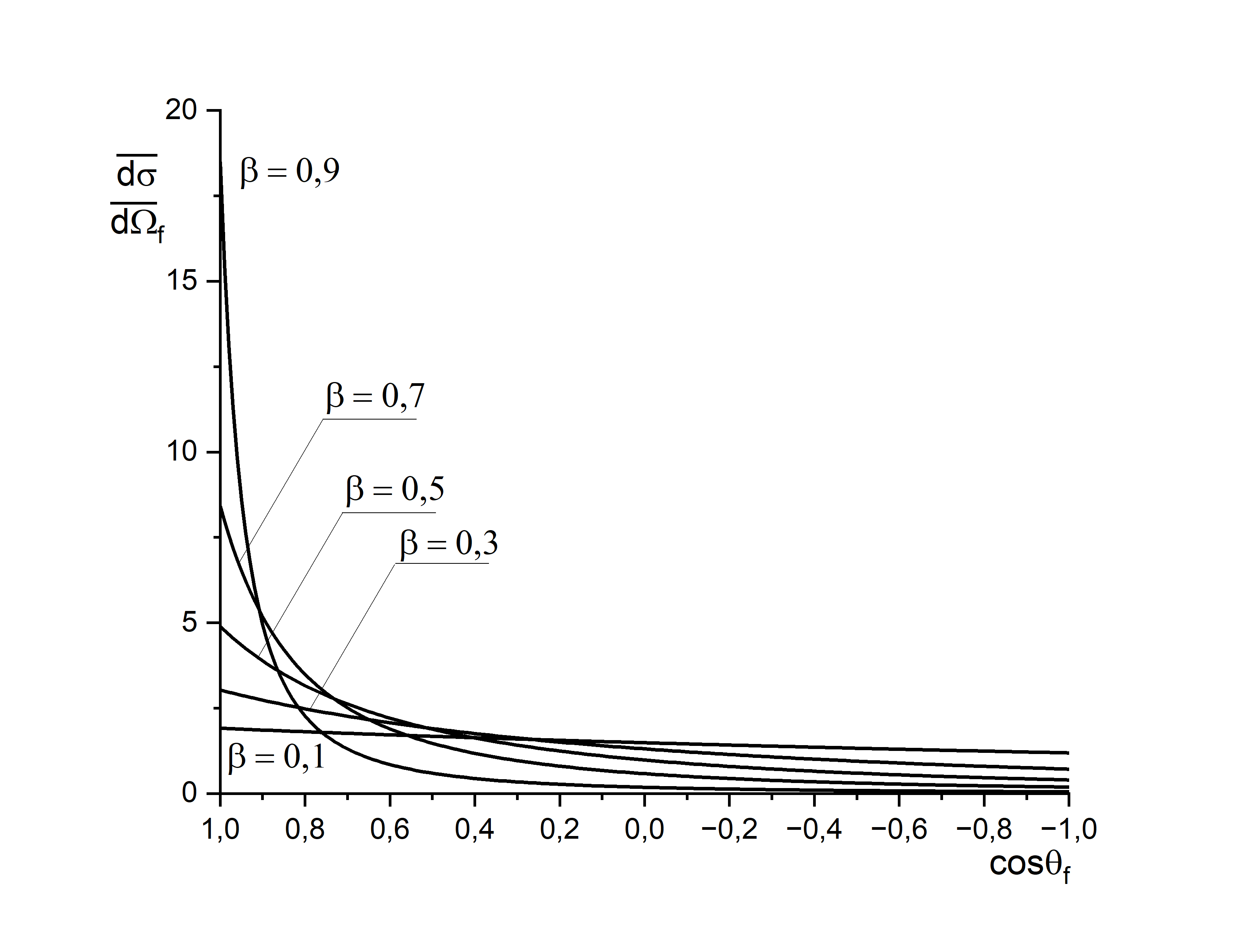}
\caption{\large Left panel: Differential cross section (6.1)
depending on the scattering angle $\theta_f$ for different values
of $\beta$. The incident photon energy is $\omega_i=150$ keV, the
angle is $\theta_i=180^\circ$ (head-on collision). Right panel:
Averaged differential cross section (7), calculated taking into
account Eqs. (6.1) and (3), $\theta_i=90^\circ$}
\end{figure}

This effect can be partly explained as follows. We expand cross
section (7)  into a series in Legendre polynomials
$P_l(\cos\theta_{fi})$ and use the well-known formula
$$
P_l(\cos\theta_{fi})=P_l(\cos\theta_f)P_l(\cos\theta_i) +2\sum_{m=1}^l \frac{(l-m)!}{(l+m)!}\ P_l^m(\cos\theta_f)P_l^m(\cos\theta_i)\cos m\Phi.
$$
Integration with respect to the angle $\Phi$ removes the sum, and
only the first term remains in the expansion. If the angle
$\theta_i\sim 180^\circ$, then $P_l(-1)=\pm 1$ depending on  the
parity of $l$. If the angle $\theta_i\sim 90^\circ$, then the odd
polynomials $P_l(0)=0$, while the even polynomials are not equal
to 1. From a mathematical point of view, the difference in cross
sections is explainable, but we still have no physical
explanation.

A completely surprising situation with the cross section arises
when $\theta_i=\theta_f=0$, i.e. the initial photons fly
collinearly the electron beam and interact with it. The frequency
of photons in this case does not change, but cross section (6.1)
for forward scattering is equal to
$$
\frac{d\sigma}{d\Omega_f}=r^2_0\frac{1+\beta}{1-\beta},
$$
and grows infinitely in the ultrarelativistic case. This behavior
is due to tending to zero with increasing energy of the
denominator of the electron propagator between two vertices of the
Feynman diagram, when the momenta of the photon and electron are
parallel.

For a more detailed study of this interesting phenomenon, we
plotted the dependence of differential cross section (6) on the
angle of incidence of the photon $\theta_i$ on a moving electron
for a fixed scattering angle $\theta_f=0$, which is presented in
Fig. 3. The dependence on the angle $\Phi$ is absent in this case.
In addition, in Fig. 3 we have presented, for convenience, graphs
of frequency ratio $\omega_f/\omega_i$. From Fig. 3 it follows,
that  the cross section is practically constant at low electron
velocities $\beta$ and the change in frequency is small. As
$\beta$ grows, there appears a maximum  in the cross section,
which shifts  to smaller angles of incidence of the photon with
increasing $\beta$. At the same time,   the change in frequency
close to the maximum is, of course, significantly smaller than at
the angle $\theta_i=180^\circ$.

This phenomenon is an unexpected and curious one, and has no
direct relation to the inverse Compton effect with its goal to
maximize the frequency of the final photon. However, we have found
it to be interesting and tried to explain by carrying out a
preliminary investigation of the derivative of  cross section
(6.1) with respect to $\cos\theta_i$  depending on $\beta$. It
turned out that the maximum is already formed at $\beta\approx
0.2$ for $\cos\theta_i\sim -1$, i.e. for the geometry of
backscattering. As $\beta$ increases,  the value of
$\cos\theta_i^{max}$ corresponding to the maximum of the cross
section  grows quite sharply, crosses zero at $\beta\approx 0.4$
and tends to one at asymptotic (ultrarelativistic) electron
energies, and in this case
$\cos\theta_i^{max}(\beta)\approx\beta$. We have obtained this
result for the photon energy $\omega_i = 150$ keV. A more complete
study of this curious phenomenon requires consideration of other
cases.

\begin{figure}[ht!]
\centering
\includegraphics[scale=0.35]{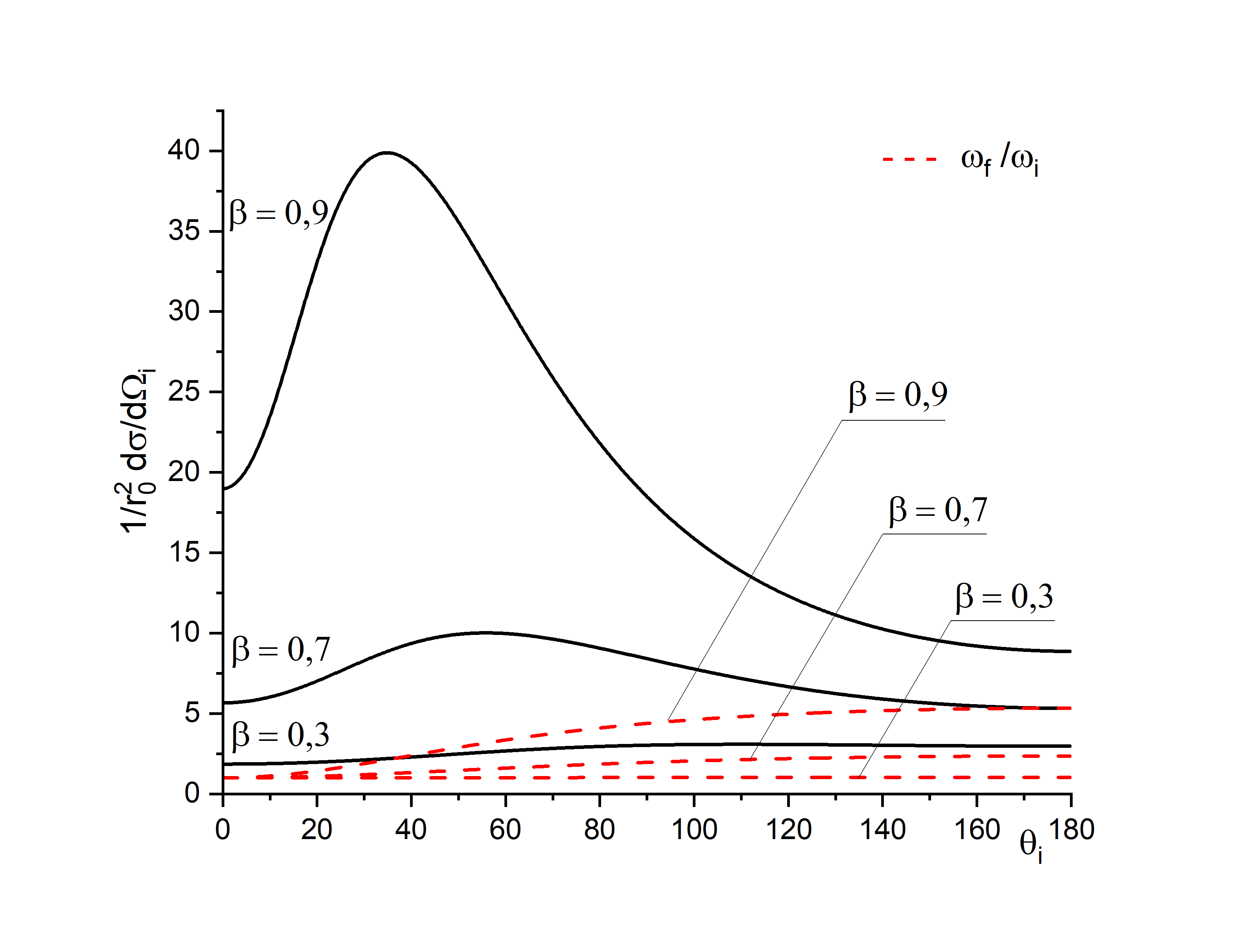}
\caption{\large Differential cross section (6.1) calculated with
taking into account formulas (6.2) and (3) depending on the angle
of incidence of the photon $\theta_i$ for different $\beta$. The
scattering angle is fixed, $\theta_f=0^\circ$. The incident photon
energy is $\omega_i=150$ keV. Black (solid) lines show cross
sections. For convenience, the red (dashed) lines show the ratio
$\omega_f/\omega_i$.}
\end{figure}

\subsection*{\small POLARISATION EFFECTS}

Let us now consider the influence of the polarizations of
colliding particles on the cross section. In describing
polarization effects, we use again the formulas from monograph
\cite{AB}. We consider first the most common case of scattering of
polarized photons at unpolarized electrons, where the
polarizations of the final particles are not measured. The
scattering plane of photons is given by the $(z,y)$ axes, and the
$x$ axis is orthogonal to this plane. The degree of linear
polarization is characterized by the Stokes parameter (the index
$i$ refers to the initial photon or electron) $\xi_3^{(i)}=\cos
2\alpha$, where the angle $\alpha$ determines the polarization of
the photon with respect to the $x$ axis. The photons polarized
transverse to the scattering plane are characterized by the angle
$\alpha=0$, in the scattering plane $\alpha=\pi/2$. The angle
$\alpha=\pi/4$ corresponds to the absence of linear polarization.

In this case the cross section  is also given by Eq. (6.1), where,
however,
$$
U_0(\xi_3^{(i)})=\left[4\left(\frac{1}{\kappa_1}+\frac{1}{\kappa_2}\right)^2
- 4\left(\frac{1}{\kappa_1}+\frac{1}{\kappa_2}\right)\right](1-
\xi_3^{(i)})- \left(\frac {\kappa_1}{\kappa_2}+\frac
{\kappa_2}{\kappa_1}\right),\eqno (8)
$$

\begin{figure}[ht!]
\centering
\includegraphics[scale=0.35]{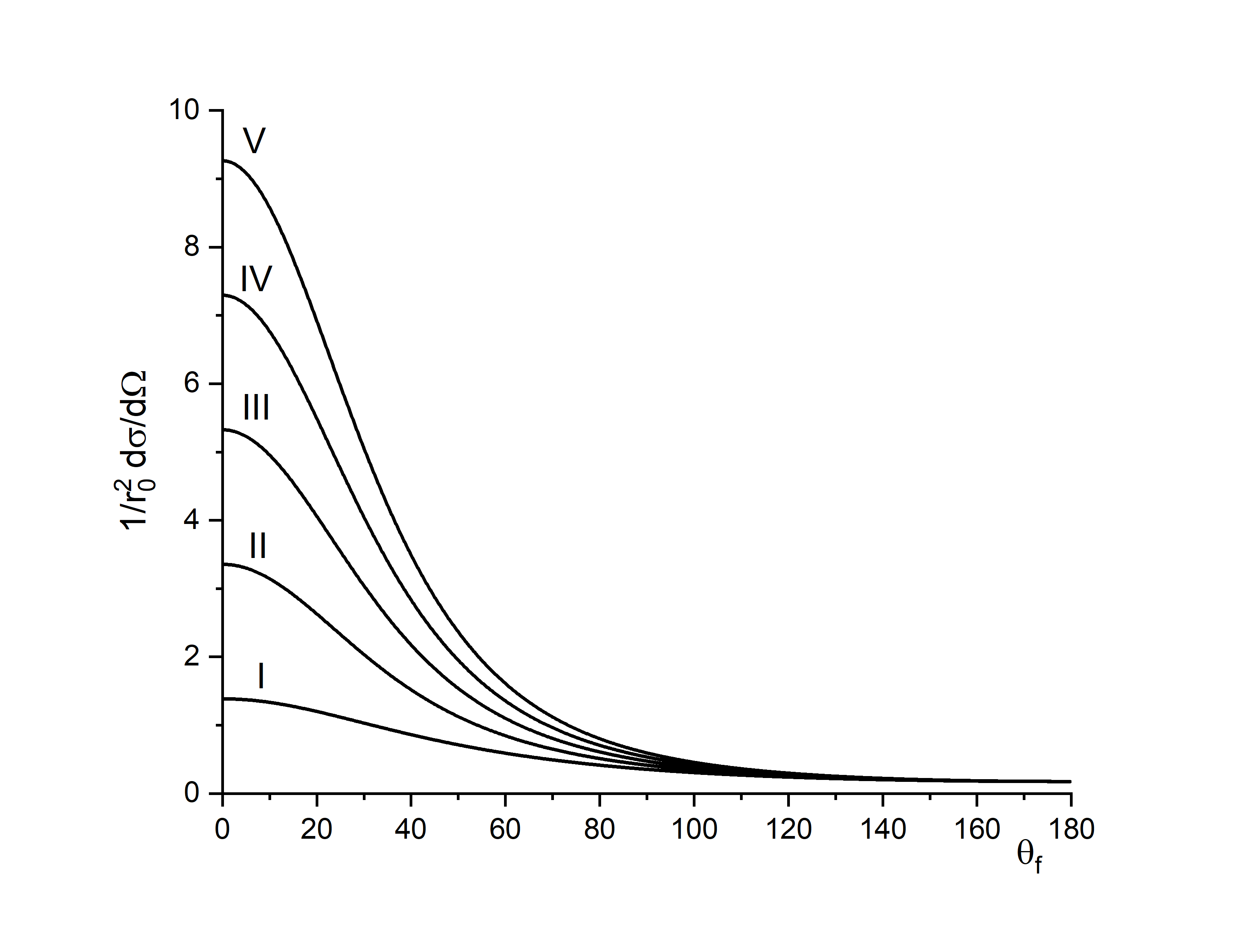}
\caption{\large Differential cross section (6.1) taking into
account (8) depending on the scattering angle $\theta_f$ for
different values of $\xi_3^{(i)}$. $I: \xi_3^{(i)}=1$; $II:
\xi_3^{(i)}=0.5$; $III: \xi_3^{(i)}=0$; $IV: \xi_3^{(i)}=-0.5$;
$V: \xi_3^{(i)}=-1$ . $\beta=0.7$.  The incident photon energy is
$\omega_i=150$ keV.}
\end{figure}

Examples of such cross sections are shown in Fig. 4. From the
figure it follows that the largest cross section is achieved, when
the photon is linearly polarized in the scattering plane, and, in
this case, the cross section is much larger than for the other
polarizations.

Next, we consider the scattering of a partially polarized photon
by a partially polarized electron. The covariant form of the cross
section in the approximation of interest to us looks as follows
(since we do not consider the polarizations of the scattered
particles, the cross section is multiplied by the factor of four):
$$
\frac{d\sigma}{d\Omega_f}=2r^2_0 \left(\frac{\omega_f}{mc^2\kappa_1}\right)^2
[U_0(\xi_3^{(i)}) +(f_2)_\mu \xi^{(i)}_2 s_\mu^{(i)} ] \eqno (9)
$$
The first term is  presented explicitly in  formula (8) and has
been considered above, the second term describes the scattering of
partially circularly polarized photons by polarized electrons. The
4-vector  $f_2$ entering it looks like:
$$
(f_2)_\mu=\frac{2}{mc}\left(\frac{1}{\kappa_1}+\frac{1}{\kappa_2}\right)\left[ k_{i\mu} +
k_{f\mu} -2\left(\frac{1}{\kappa_1}+\frac{1}{\kappa_2}\right)\ k_{i\mu}\right], \eqno (10)
$$
where $k_\mu =(\omega/c,\vec k)$.  It is necessary to remark that
formula (10) is similar to formula (26.7.5) in monograph
\cite{AB}, which is misspelled and differs from formula (10) in
the common sign that is due to the use of archaic notations in
this monograph for the coordinates in Minkowski space with
imaginary time component. In later editions of the monograph, a
consideration of polarization effects involving a moving electron
is absent at all, as well as in the most publications that we have
looked through.

The Stokes parameter $\xi^{(i)}_2$ describes a partially circular
polarization, and the components of the  spin 4-vector of a moving
electrons have the form
$$
s_0=\frac{1}{mc}(\vec p\cdot\vec\zeta), \quad \vec s=\vec\zeta + \vec p \frac{(\vec p\cdot\vec\zeta) }{mc(\sqrt{p^2+(mc)^2} +mc)}. \eqno (11)
$$
This formula gives the spin 4-vector $s$ in terms of the
3-dimensional polarization vector $\vec\zeta$ in the rest frame,
which is convenient to be used for the parametrization of the spin
vector.

\begin{figure}[ht!]
\centering
\includegraphics[scale=0.35]{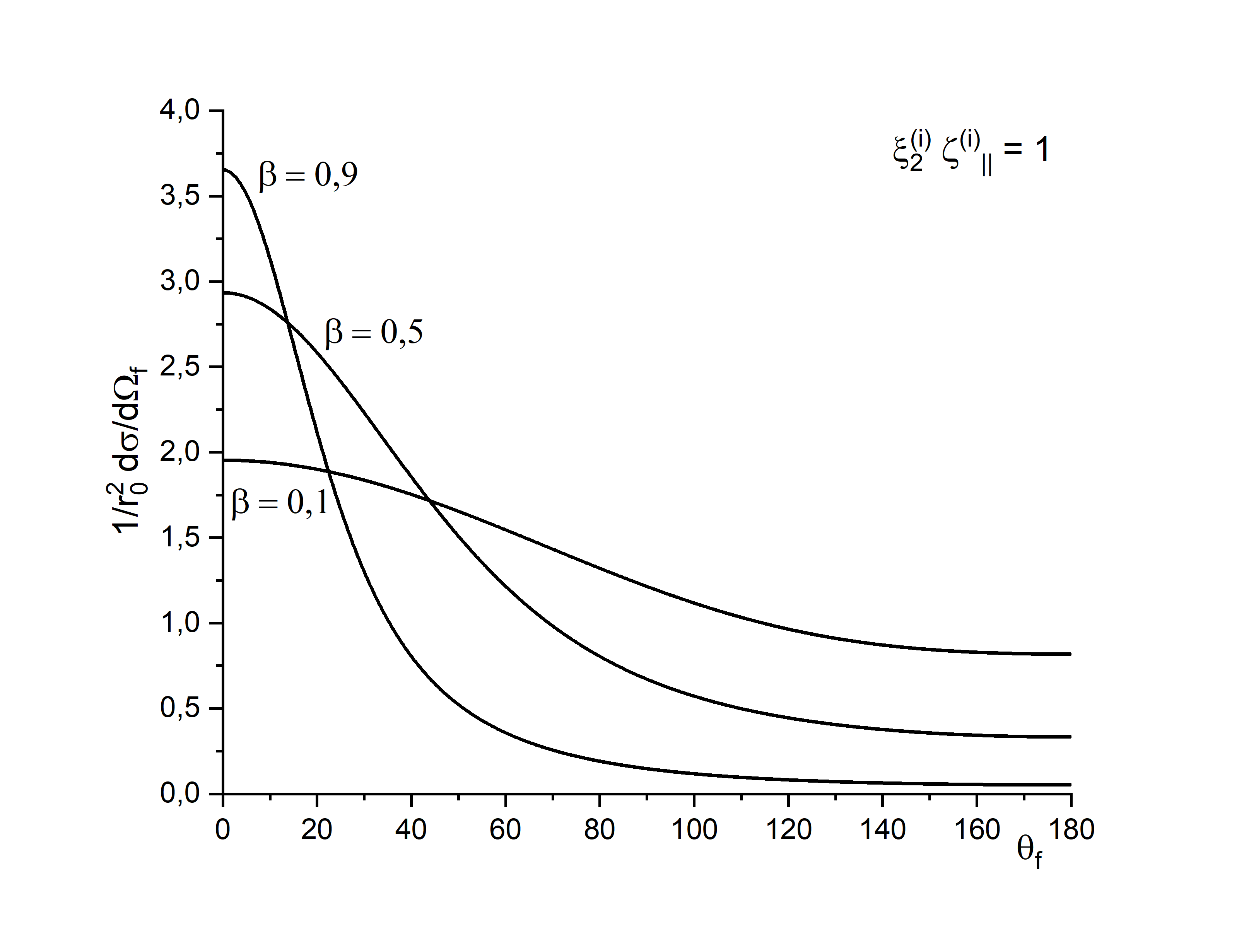}\includegraphics[scale=0.35]{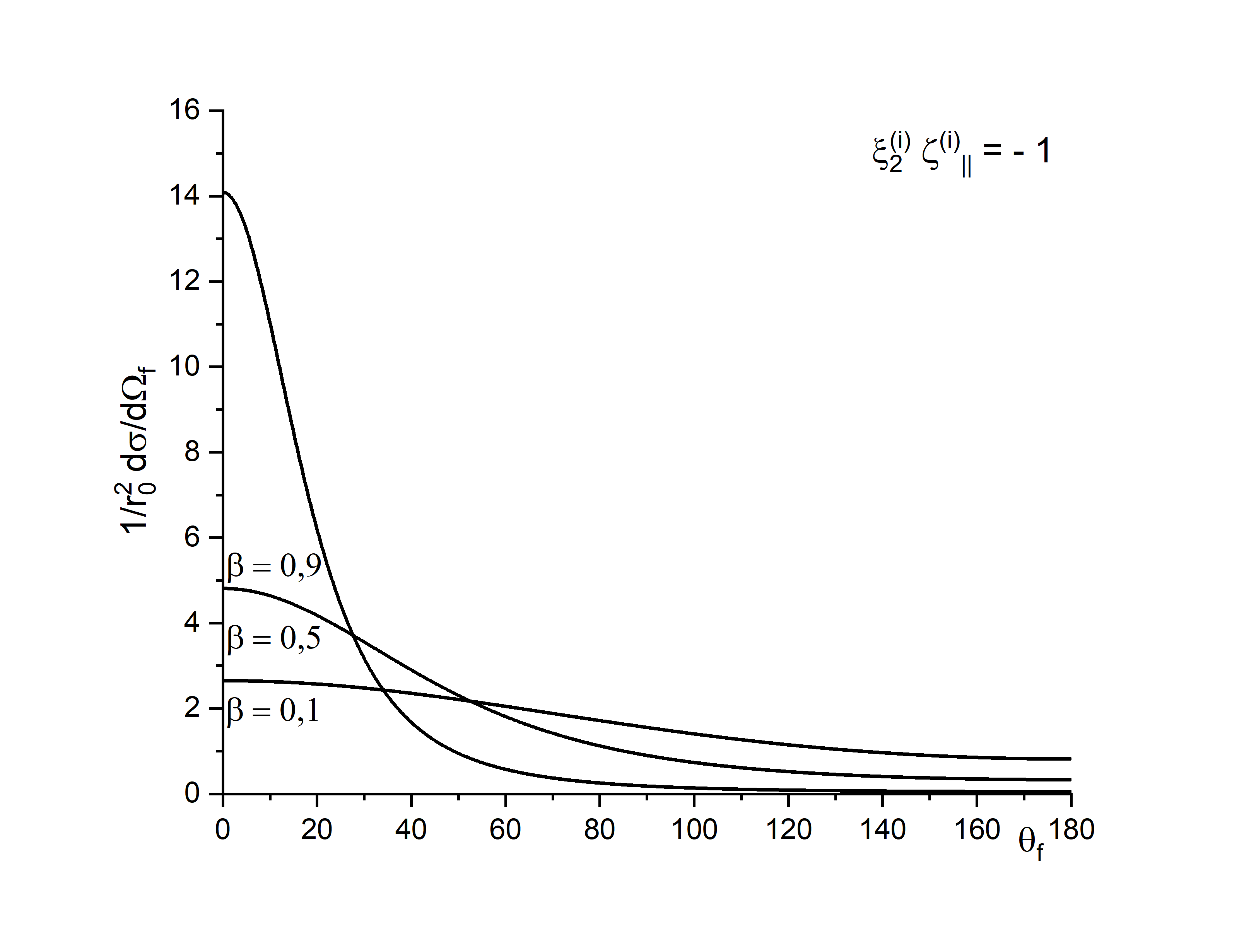}
\caption{\large Differential cross section (9) depending on the
scattering angle $\theta_f$ for different values of the helicities
of colliding particles for $\xi_3^{(i)}=0$ and several values of
$\beta$. The incident photon energy is $\omega_i=150$ keV, its
angle of incidence is $\theta_i=180^{\circ}$. Left panel: $\quad
\xi_2^{(i)}\cdot\zeta^{(i)}_\parallel=1$; right panel: $\quad
\xi_2^{(i)}\cdot\zeta^{(i)}_\parallel=-1$.}
\end{figure}

The general formula for the third term in (9) turns out to be
quite cumbersome, so we will separately consider the longitudinal
and transverse polarizations of the electron. For the transverse
polarization
$$
s_0^{(i)} = 0, \quad \vec s^{(i)} = \vec \zeta^{(i)}_\perp, \quad
(\vec p_i\cdot\vec\zeta^{(i)}_\perp) = 0, \quad |\vec \zeta^{(i)}_\perp|
\leq 1,
$$
and for the longitudinal one
$$
s_0^{(i)} = \frac{|\vec p_i|}{mc}\,\zeta^{(i)}_\parallel , \quad \vec s^{(i)} = \frac{\vec p_i
E_i}{|\vec p_i|mc^2}\, \zeta^{(i)}_\parallel, \quad  |\zeta^{(i)}_\parallel |\leq 1.
$$
Besides that,
$$
(\vec k_{i}\cdot\vec p_i) = k_{i} p_i\cos\theta_i, \quad (\vec k_{f}\cdot\vec p_i) = k_{f} p_i\cos\theta_f,
$$
$$
(\vec k_{i}\cdot\vec\zeta^{(i)}_\perp)= \frac1c\ \omega_{i}\ \zeta^{(i)}_\perp\sin\theta_i, \quad (\vec k_{f}\cdot\vec\zeta^{(i)}_\perp)= \frac1c\ \omega_{f}\ \zeta^{(i)}_\perp\sin\theta_f,
$$
The last two formulas are written for the case, where the electron
spin vector lies in the scattering plane, and $\zeta^{(i)}_\perp$
can take both positive and negative values.

As a result, for the transverse polarization of the electron  in
the scattering plane, after some transformations, the general
formula can be rewritten as follows:
$$
((f_2)_\mu \xi^{(i)}_2 s_\mu^{(i)})_\perp =
-\xi^{(i)}_2\zeta^{(i)}_\perp\left(\frac{2\omega_i}{mc^2}\right)\left(\frac{1}{\kappa_1}+\frac{1}{\kappa_2}\right)\ \times
$$
$$
\left[
\sin\theta_i\left(1-2\left(\frac{1}{\kappa_1}+\frac{1}{\kappa_2}\right)\right)
 + \frac{\omega_f}{\omega_i}\sin\theta_f\right], \eqno (12.1)
$$
and for the longitudinal polarization
$$
((f_2)_\mu \xi^{(i)}_2 s_\mu^{(i)})_\parallel =
\xi^{(i)}_2\zeta^{(i)}_\parallel \left(\frac{2\omega_i}{mc^2\sqrt{1-\beta^2}}\right)\left(\frac{1}{\kappa_1}+\frac{1}{\kappa_2}\right)\ \times
$$
$$
\left[(\beta -\cos\theta_i)\left(1
-2\left(\frac{1}{\kappa_1}+\frac{1}{\kappa_2}\right)\right)
 + \frac{\omega_f}{\omega_i}(\beta -\cos\theta_f) \right].  \eqno (12.2)
$$
From Eq. (12.1), in particular, it follows that if $\theta_i=\pi$
and $\theta_f=0$, then the transverse polarization of the electron
does not contribute to the cross section at all.

\begin{figure}[ht!]
\centering
\includegraphics[scale=0.35]{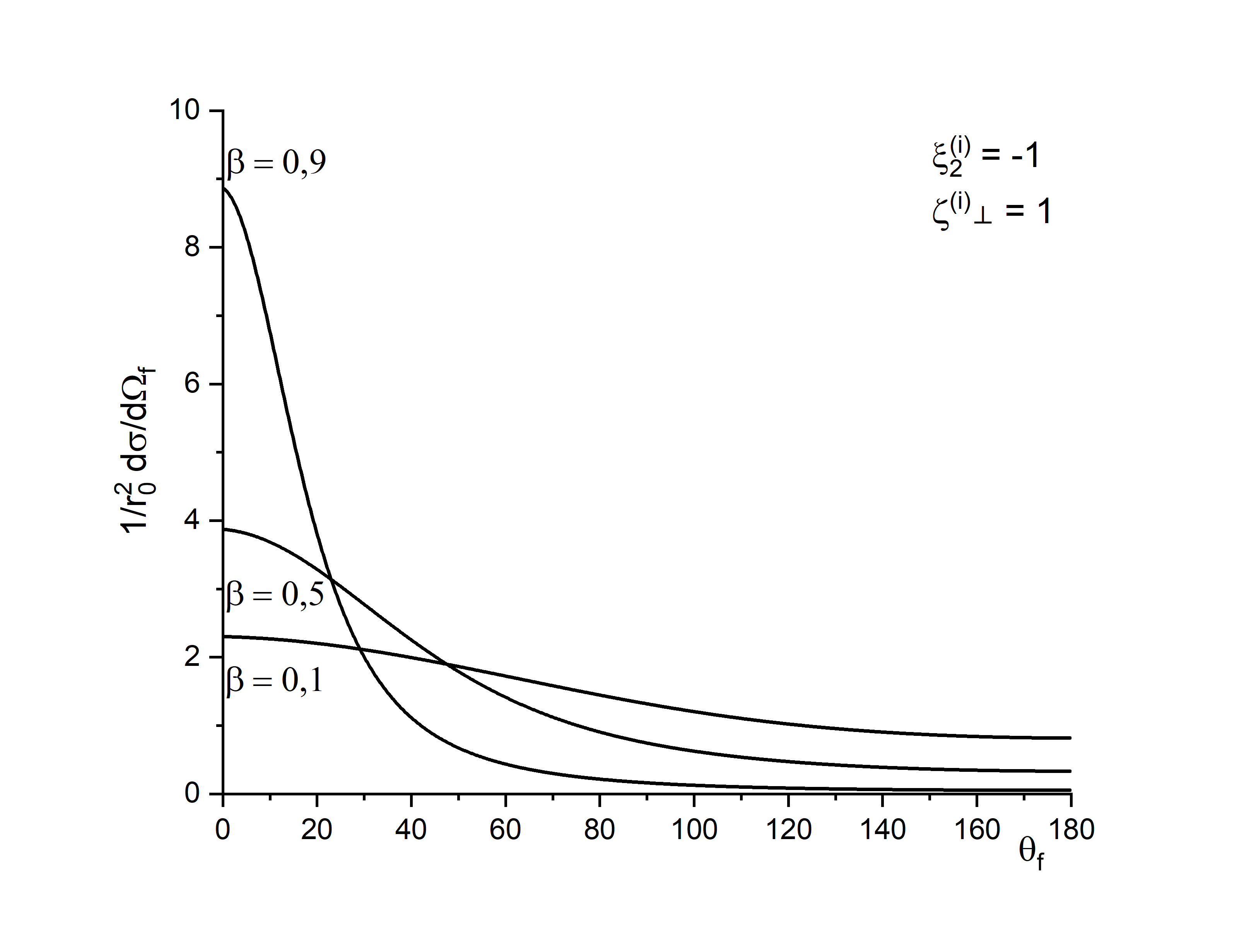}
\caption{\large Differential  cross section (9) depending on the
scattering angle $\theta_f$ for different values of the
polarizations of colliding particles: $\xi_3^{(i)}=0, \quad
\xi_2^{(i)}=-1, \quad \zeta^{(i)}_\perp=1$, and for several values
of $\beta$. The incident photon energy is $\omega_i=150$ keV, its
angle of incidence is $\theta_i=180^{\circ}$.}
\end{figure}

Figs. 5,6 present examples of calculating differential cross
section (9), where the linear polarization of the photon
$\xi^{(i)}_3=0$. In Fig. 5 one sees the cross sections for the
same and opposite helicities of photons and electrons. Because the
polarization parameters enter cross section (12.2) as a product,
then the cases $\xi^{(i)}_2\cdot\zeta^{(i)}_\parallel=\pm 1$
correspond to the same (opposite) helicities. Comparing Fig. 5 and
2 (left panel), we see that the electron helicity has a noticeable
effect on the value of the cross section, and the maximum
contribution to the cross section for the forward photon
scattering is given by the polarization term with opposite photon
and electron helicities. On the contrary, the same helicities of
the particles noticeably reduce the cross sections.

As for the transverse polarization of the electron (Fig. 6), the
contribution of the term $((f_2)_\mu \xi^{(i)}_2
s_\mu^{(i)})_\perp$  to cross section (9) is very small, and it
changes very little the cross section compared to the unpolarized
colliding particles shown in the left panel of Fig. 2. As a
consequence, the sign of the photon helicity $\xi^{(i)}_2$ does
not matter.

\section*{\large CONCLUSION}

We have considered some kinematic conditions, including the
polarizations of the colliding electron and photon, at which the
maximum values of the differential cross section are reached at
given electron and photon energies. It has been found that the
cross section increases noticeably in the case of the electron
spin being directed along its velocity  and the left photon
helicity, or vice versa.

Moreover, an unexpected effect of an increase  in the cross
section has been found for angles of incidence of  photons on an
electron beam different from $\pi$, while the final photons move
along the beam. It has been shown that, as the angle of incidence
of the photons decreases, the cross section increases and reaches
a peak, the value of which increases with the growth of the
electron energy, although such kinematics is not directly related
to the classical inverse Compton effect.

\subsection*{\small ACKNOWLEDGEMENTS}

The work was carried out with  financial support of the Ministry
of Science and higher education of the Russian Federation, project
No.075-15-2021-1353.

\end{document}